\pdfoutput=1

\documentclass[showpacs,aps,prl,twocolumn,superscriptaddress]{revtex4}
\usepackage{graphicx} 
\usepackage{dcolumn}
\usepackage{bm}
\usepackage{amssymb,amsmath}
\usepackage{epstopdf}
\usepackage[T1]{fontenc}
\usepackage[latin9]{inputenc}
\usepackage{color}
\usepackage{float}
\usepackage{esint}
\usepackage{graphics}
\makeatletter
\@ifundefined{textcolor}{}
{%
 \definecolor{BLACK}{gray}{0}
 \definecolor{WHITE}{gray}{1}
 \definecolor{RED}{rgb}{1,0,0}
 \definecolor{GREEN}{rgb}{0,1,0}
 \definecolor{BLUE}{rgb}{0,0,1}
 \definecolor{CYAN}{cmyk}{1,0,0,0}
 \definecolor{MAGENTA}{cmyk}{0,1,0,0}
 \definecolor{YELLOW}{cmyk}{0,0,1,0}
}

\begin{document}

\title{Single-Shot Terahertz Time-Domain Spectroscopy in Pulsed High Magnetic Fields}
\normalsize

\author{G.~Timothy Noe II}
\thanks{Author to whom correspondence should be addressed}
\email[]{TimNoe@rice.edu}
\affiliation{Department of Electrical and Computer Engineering, Rice University, Houston, Texas 77005, USA}

\author{Ikufumi Katayama}
\affiliation{Department of Physics, Graduate School of Engineering, Yokohama National University, Yokohama 240-8501, Japan}

\author{Fumiya Katsutani}
\author{James~J.~Allred}
\author{Jeffrey~A.~Horowitz}
\author{David~M.~Sullivan}
\author{Qi Zhang}
\affiliation{Department of Electrical and Computer Engineering, Rice University, Houston, Texas 77005, USA}

\author{Fumiya Sekiguchi}
\affiliation{Department of Physics, The University of Tokyo, Tokyo, 113-0033, Japan}

\author{Gary L.~Woods}
\affiliation{Department of Electrical and Computer Engineering, Rice University, Houston, Texas 77005, USA}

\author{Matthias C.~Hoffmann}
\affiliation{SLAC National Accelerator Laboratory, Menlo Park, California 94025, USA}

\author{Hiroyuki Nojiri}
\affiliation{Institute for Materials Research, Tohoku University, Sendai 980-8577, Japan}

\author{Jun Takeda}
\affiliation{Department of Physics, Graduate School of Engineering, Yokohama National University, Yokohama 240-8501, Japan}

\author{Junichiro Kono}
\affiliation{Department of Electrical and Computer Engineering, Rice University, Houston, Texas 77005, USA}
\affiliation{Department of Physics and Astronomy, Rice University, Houston, Texas 77005, USA}
\affiliation{Department of Materials Science and NanoEngineering, Rice University, Houston, Texas 77005, USA}

\date{\today}

\begin{abstract}
We have developed a single-shot terahertz time-domain spectrometer to perform optical-pump/terahertz-probe experiments in pulsed, high magnetic fields up to 30\,T. The single-shot detection scheme for measuring a terahertz waveform incorporates a reflective echelon to create time-delayed beamlets across the intensity profile of the optical gate beam before it spatially and temporally overlaps with the terahertz radiation in a ZnTe detection crystal. After imaging the gate beam onto a camera, we can retrieve the terahertz time-domain waveform by analyzing the resulting image. To demonstrate the utility of our technique, we measured cyclotron resonance absorption of optically excited carriers in the terahertz frequency range in intrinsic silicon at high magnetic fields, with results that agree well with published values.
\end{abstract}

\maketitle


\section{Introduction}

Terahertz time-domain spectroscopy (THz-TDS) is ideally suited for the study of coherent low-energy dynamics in condensed matter, determining the complex conductivity of the sample with a high signal-to-noise ratio (SNR)~\cite{Nuss98,Jepsen11,Ulbricht11,Lloyd-Hughes14}.  In the presence of a strong external magnetic field, a variety of elementary and collective THz excitations occur associated with spin and orbital quantization. Probing these excitations through THz-TDS in high magnetic fields~\cite{Some94,Huggard97,Wang07,Lloyd-Hughes08,Wang10,Ikebe10,Arikawa11,Valdes-Aguilar12,Scalari12,Zhang14,Kamaraju15,Curtis16,Zhang16} can provide valuable insight into the quantum coherent and nonequilibrium dynamics of interacting electrons, while potentially leading to many-body strategies for quantum-based technologies. However, performing THz-TDS experiments in high magnetic fields above 10~T remains a formidable technical challenge. One obstacle has been coupling broadband THz radiation, often difficult to collimate, into and out of the sample space of the, generally, large magnet systems required for generating high magnetic fields. An attempt to overcome that challenge in a direct current (DC) magnet involved placing fiber-coupled THz emitter and receiver packages in the magnet bore, but this led to difficulty aligning and calibrating these components after cooling down the system and applying the magnetic field~\cite{Crooker02}. Ultrafast pump-probe spectroscopy has been demonstrated up to 25~T in the recently developed Split Florida-Helix~\cite{Curtis14}. Designed with ultrafast spectroscopy (including THz-TDS) in mind, this magnet has direct optical access via large window ports to reduce dispersion/pulse broadening due to optical fibers. However, the large size of the magnet, $\sim$1~m in diameter, should make aligning THz radiation through the sample space of the system a significant challenge, as broadband THz radiation is difficult to collimate and focus over long distances.

For pulsed magnets, which can generate a higher magnetic field with a smaller magnet and cryostat size than DC magnets~\cite{Herlach03}, the challenge is the speed with which the THz waveform can be measured. Step-scan methods to ultrafast time-domain measurements with pulsed magnets have been reported, where one magnet shot is required for each gate pulse delay time for both THz-TDS~\cite{Molter12} and time-resolved photoluminescence~\cite{Noe13}. For our magnet system, because of the wait time between magnet shots, on the order of several minutes for shots up to 30~T~\cite{Noe13}, this method proves to be unreasonably time-consuming for recording the gate signal at many time delays. To date, reported rapid scanning methods for THz-TDS in pulsed magnets include using a rotating delay line for measurements up to 12~T with a 40~T magnet~\cite{Molter10}, an electronically controlled optical sampling (ECOPS) method up to 2.5~T with a 30~T magnet~\cite{Noe14}, and, most recently, an asynchronous optical sampling (ASOPS) method up to 13~T with a 31~T magnet~\cite{Spencer16} to change the relative timing between the pump and gate pulse for each consecutive pair of laser pulses. This leads to a mapping of the pump/probe delay time to the real measurement time. For instance, in Ref.~\cite{Noe14}, 15~ps of time delay corresponds to 150~$\mu$s of measurement time, which provides slightly less than 1\% of magnetic field variation at the peak of the magnetic field pulse that varies on a timescale of milliseconds. For shorter-duration magnetic field pulses, for instance $\mu$s pulses in a single-turn coil system, required for ultrahigh magnetic field strengths~\cite{Herlach03}, these methods are simply not fast enough, i.e., the magnetic field variation significantly complicates data extraction.

Single-shot methods can provide probe time delay information after a pump pulse for a range of delays, typically up to 10s of ps, utilizing only a single laser pulse. This can be achieved by encoding time delay information on a spectrally chirped gate pulse or by utilizing a specially designed `stair step' optic to intentionally control the delay of portions of the pulse front. Spectral encoding techniques can induce distortions in the signal if the bandwidth of the gate pulse is insufficient as compared to that of the THz pulse~\cite{Yellampalle07}.  Recently, Teo {\it et al}.\ reviewed existing single-shot THz-TDS techniques~\cite{Teo15} for multidimensional THz spectroscopy, suggesting the dual transmissive echelon technique to be an excellent method. The authors also mention a transition to using a large reflective echelon~\cite{Minami13} instead of the transmissive echelons for greater frequency resolution and larger time windows; it also simplifies data analysis processes because it only uses one dimension for the time axis. Single-shot THz-TDS has also been used in optical-pump/THz-probe (OPTP) spectroscopy experiments, determining the complex dielectric constant of photogenerated carriers in intrinsic Si with high SNRs~\cite{Minami15}.

Here, we have developed a single-shot THz time-domain spectrometer utilizing a reflective echelon mirror to measure the THz response of materials at magnetic fields up to 30~T, sample temperatures down to $\sim$10~K, and with the option of optical pumping. We have demonstrated the utility of the system by measuring the cyclotron resonance absorption of photoexcited carriers in bulk, intrinsic (100) silicon (Si) under these experimental conditions.

\section{Experimental Setup}

\begin{figure*}
\centering
\includegraphics[scale=1.6]{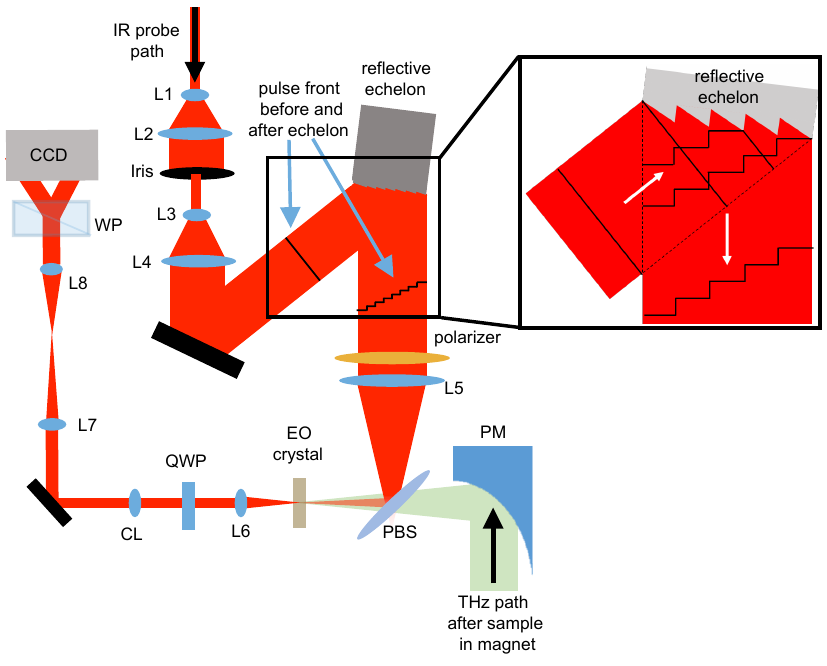}
\caption{Schematic diagram of the single-shot detection scheme. Two 10$\times$ telescopes, L1-L4, are used to expand the optical gate beam so that a relatively uniform intensity profile reflects off of the echelon mirror. After encoding time delay information onto the intensity profile of the gate beam with the echelon optic, the intensity profile at the plane of the echelon surface is imaged onto the electro-optic sampling crystal after reflecting off of a pellicle beam splitter, PBS, overlapping with the THz beam and finally onto a CCD camera with image relay optics, L5-L8. A Wollaston prism, WP, is used to separate the two orthogonal polarizations of the elliptically polarized gate beam after a quarter-wave plate, QWP, and a cylindrical lens, CL, is used to focus the beam in one direction so that both polarizations can be measured with a single CCD camera. Note that the Wollaston prism and the resulting separated beams are shown rotated 90$^{\circ}$ with respect to the actual orientation.}
\label{ill:Setup}
\end{figure*}

\begin{figure*}
\centering
\includegraphics[scale=0.54]{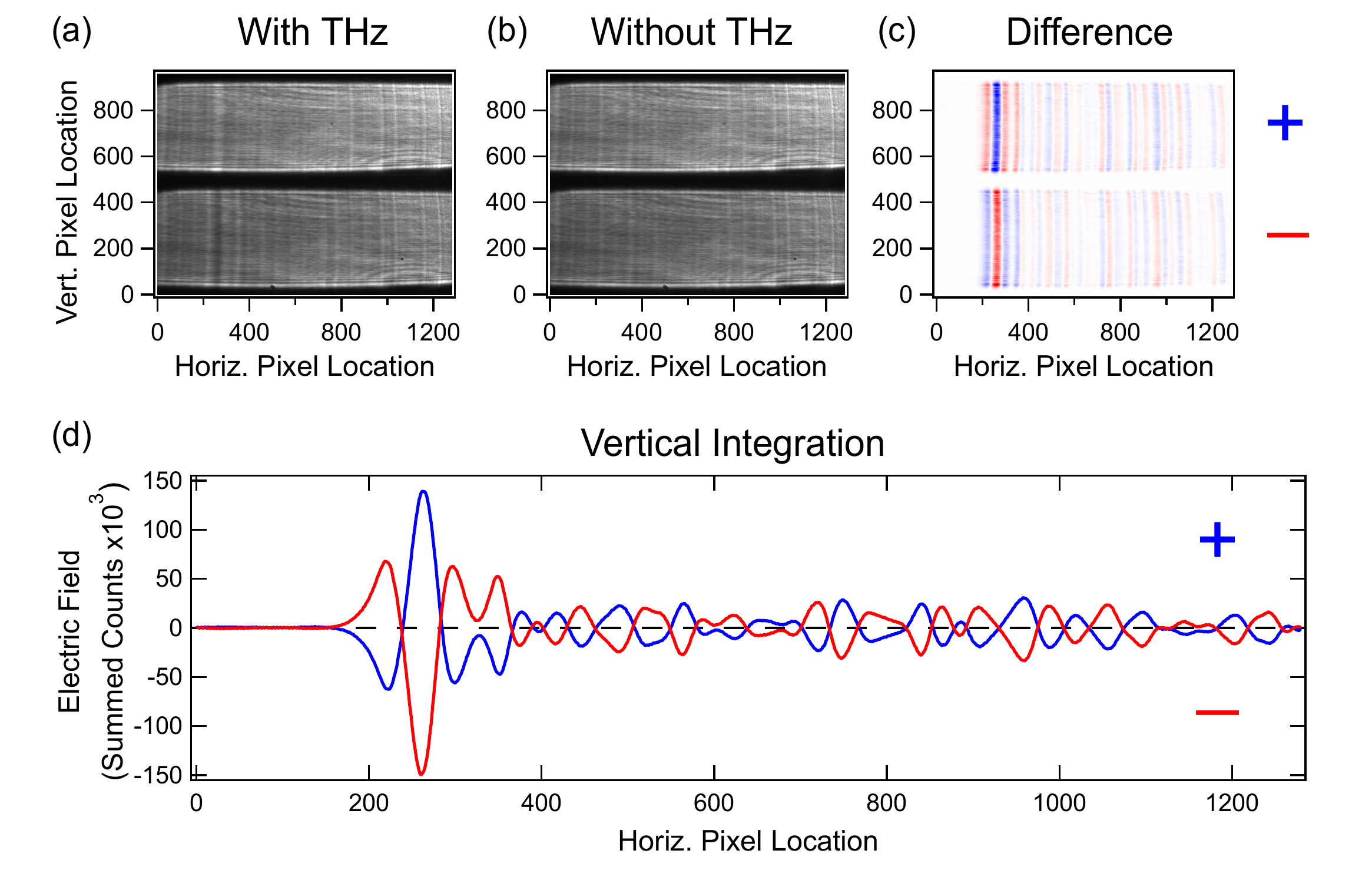}
\caption{Extracting the THz signal from the camera images. Utilizing a Wollaston prism, a cylindrical lens, and a series of relay lenses, two orthogonally polarized (denoted + and -) components of the elliptically polarized gate beam are imaged onto separate portions of the camera, \textbf{a)} and \textbf{b)}. The images of the gate beam are shown with, \textbf{a)}, and without, \textbf{b)}, the THz beam incident on the ZnTe detection crystal.  Images \textbf{a)} and \textbf{b)} are the result of averaging 100 images of individual laser pulses. The difference, \textbf{c)}, between image \textbf{a)} and image \textbf{b)} is calculated by simply subtracting the number of counts at each pixel. Oscillations due to water vapor absorption are clearly evident in the difference image \textbf{c)}. After vertically summing the top and bottom halves of the difference image, the resultant traces for each polarization component are shown in \textbf{d)} to be very close to mirror images of one another.}
\label{ill:Images}
\end{figure*}

\begin{figure*}
\centering
\includegraphics[scale=0.57]{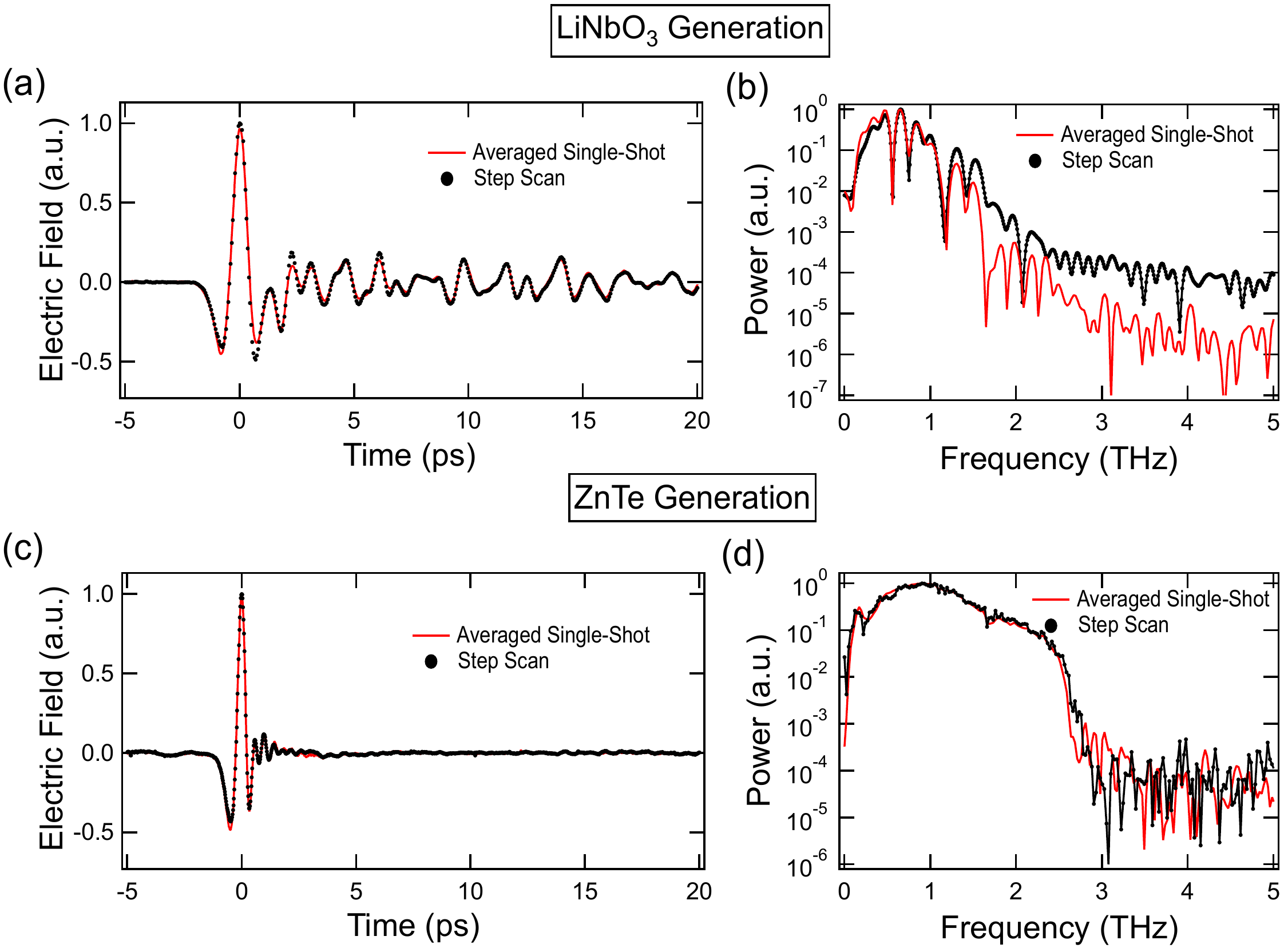}
\caption{Comparison of the single-shot technique with the step scan technique for both LiNbO$_{3}$ THz generation, \textbf{a)} and \textbf{b)}, and ZnTe THz generation, \textbf{c)} and \textbf{d)}. These measurements were performed without the cryostats in place. For the LiNbO$_{3}$ result, \textbf{a)}, the THz beam path was not purged of water vapor whereas the THz beam path for the ZnTe result, \textbf{c)}, was in a dry nitrogen purged environment. All traces are normalized to their peak value.}
\label{ill:Comparisons}
\end{figure*}

We use an amplified Ti:sapphire laser system (Clark-MXR, Inc. CPA-2001) to generate laser pulses centered at 775~nm with 1~kHz repetition rate, 150~fs pulse duration, and 1~mJ pulse energy. A beam splitter is placed in the path of the beam to reflect 20$\%$ of the power for optically pumping a sample and transmit 80$\%$ of the power for the THz generation and detection portions. Two mirrors are placed on a 500~mm linear stage in the path of the portion of the beam for THz generation and detection to provide a maximum optical delay up to 3.33~ns 
between the optical pump and the THz probe pulses. A second beam splitter is placed in the path after the delay stage to reflect 10$\%$ of the remaining power for single-shot THz detection and transmit 90$\%$ of the remaining power for THz generation. A second linear delay stage after this beam splitter is used to adjust the timing of the gate pulse with respect to the THz pulse. The THz generation section utilizes the tilted-pulse-front excitation method~\cite{Hebling08} for generating THz pulses in Mg-doped stoichiometric LiNbO$_{3}$ or the more common ZnTe generation for a greater THz bandwidth. The generated THz radiation is focused through the sample in the pulsed magnet cryostat system and then finally focused onto a 1-mm-thick (110) ZnTe electro-optic sampling crystal with an off-axis parabolic mirror. A pellicle beam splitter is used to reflect an unfocused optical pump beam towards the sample for OPTP measurements.

For the detection of THz radiation (see Fig.~\ref{ill:Setup}), a reflective echelon mirror is used to encode time delay information across the intensity profile of the near-infrared gate beam. The echelon mirror is 20~mm $\times$ 20~mm in size with 1000 steps of 20~$\mu$m width and step height of 5~$\mu$m. Upon reflection from the echelon mirror, the 150~fs gate pulse becomes a $\sim$33 ps pulse with a stair-step wavefront profile. The step height corresponds to an incremental delay of $\sim$33~fs between steps; therefore, there is an oversampling of the time delay when utilizing an initially 150~fs gate pulse. The gate beam is magnified by two 10$\times$ telescopes before the echelon mirror so that a relatively uniform portion of the intensity profile reflects off of the echelon mirror. A series of lenses are used to relay the image of the beam at the face of the echelon onto a ZnTe electro-optic sampling crystal and finally onto a silicon charge coupled device (CCD) camera. A quarter-wave plate is used to balance the intensity of the two polarization components that are separated by a Wollaston prism located immediately before the camera. A cylindrical lens is used to `squeeze' the image of both polarizations such that the images for both polarizations fit on a single CCD camera (Mightex Systems, CGE-B013-U).

Figure~\ref{ill:Images} illustrates the method to extract the THz electric field from the intensity profile of the gate beam utilizing the single-shot measurement scheme. For each image, we trigger the CCD with a transistor-transistor logic (TTL) signal from the Pockels cell timing electronics of the amplified Ti:sapphire laser and set the integration time of the CCD camera to less than~1~ms, the separation between laser pulses. Here, we capture 100 images of the gate beam with the THz generation path both blocked and unblocked. After taking 100 images and averaging them for each case, we subtract the two averaged images to leave only the difference for both polarization components of the gate beam. We then separately vertically sum the remaining counts for both the top and bottom halves of the difference image. Finally, we subtract these two traces to get the THz waveform as a function of horizontal pixel location. To calibrate the horizontal axis, we adjust the delay stage in the gate beam path to shift the THz waveform to various positions on the image and use the displacement of the stage to calibrate the time delay.

Initially, we utilized the tilted-pulse-front excitation method to generate strong THz pulses in LiNbO$_{3}$ to ensure that the SNR was sufficient to perform single-shot measurements at high magnetic fields. After successfully demonstrating the single-shot technique with LiNbO$_{3}$ generation, we switched to a ZnTe crystal for generating THz radiation to increase the frequency bandwidth of the overall spectrometer. For many semiconductors with a relatively small carrier effective mass, $m^{*}$, the frequency of cyclotron resonance can exceed the bandwidth of THz generated with LiNbO$_{3}$. Figure~\ref{ill:Comparisons} shows the measured THz electric field in the time domain and calculated power in the frequency domain for both generation schemes. Furthermore, we compare the result of the single-shot scheme with the commonly used step-scan method to measure a THz time-domain waveform. In this comparison, the result of the single-shot technique utilizes a total of 200 laser pulses (100 with the THz beam blocked and 100 with the THz beam unblocked), whereas the step-scan method uses a total of $\sim$100 pulses per time delay with $\sim$500 time delay steps, so, in total, $\sim$250 times as many pulses are used for the step-scan method to achieve a similar SNR. Based on the dimensions of the reflective echelon and the subset of the gate beam imaged onto the CCD camera, the time window for our measurement is $\sim$25~ps, which allows for a frequency resolution of $\sim$40~GHz. Overall, the agreement between the single-shot and step-scan techniques is excellent.

\begin{figure}[h]
\centering
\includegraphics[scale=0.7]{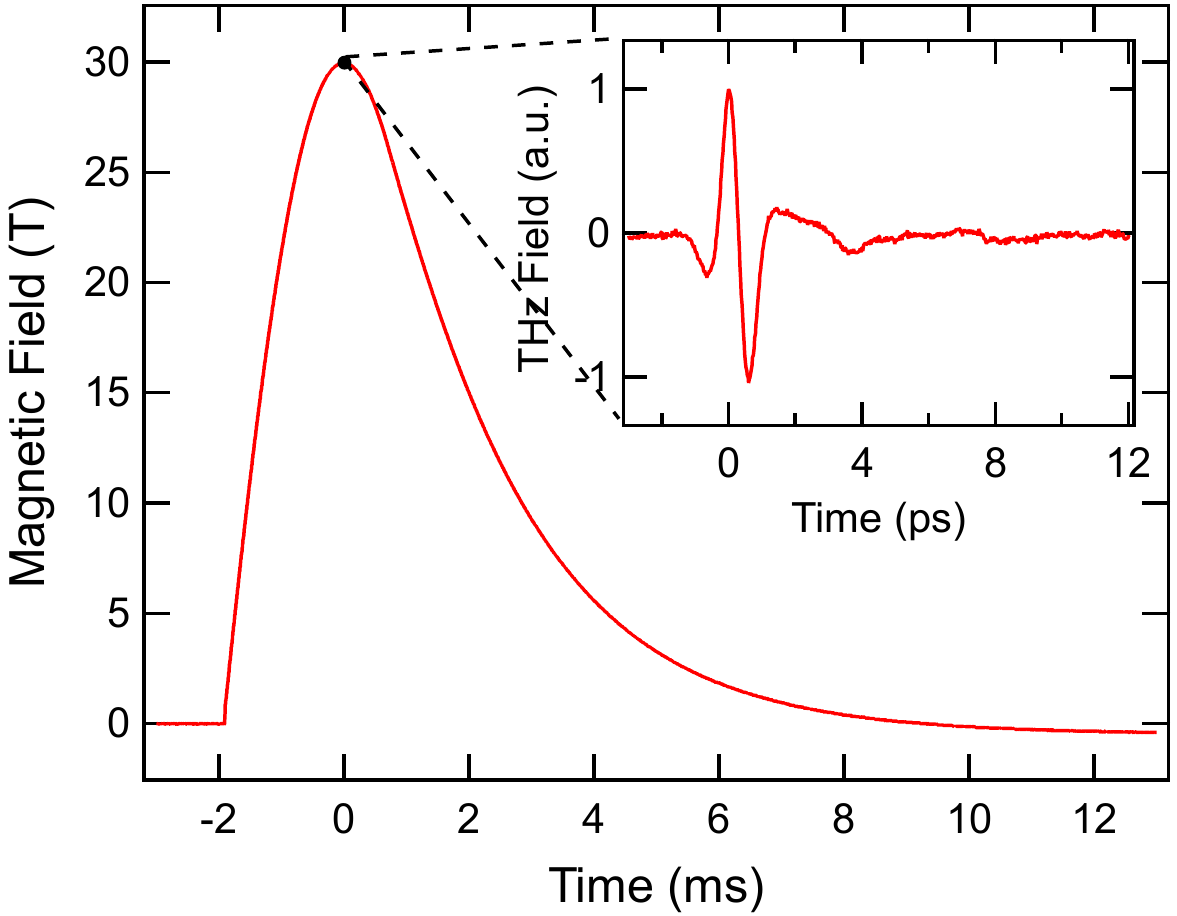}
\caption{Illustration showing the measured THz electric field after passing through a silicon sample in the cryostat system at the peak of the applied magnetic field. The THz radiation in this trace was generated in LiNbO$_{3}$, and the THz trace is a result of a single-shot measurement. The time scale of the magnetic field pulse is $\sim$10$^9$ times slower than the time scale of the THz pulse; therefore, the magnetic field variation during the interaction of the THz radiation with the sample is negligibly small.}
\label{ill:PulseDemo}
\end{figure}

\begin{figure}
\centering
\includegraphics[scale=0.62]{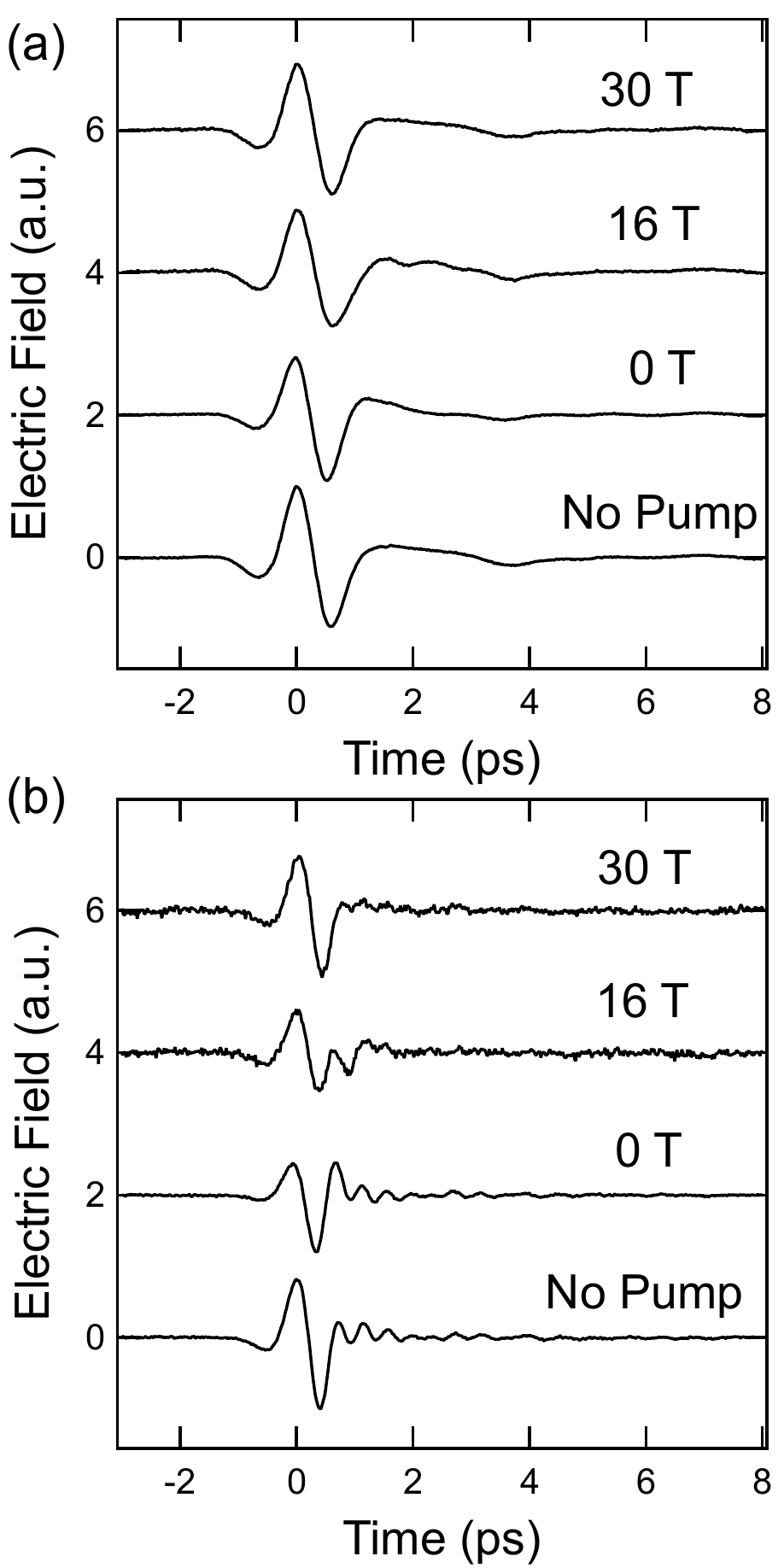}
\caption{Measured THz waveforms for both the LiNbO$_{3}$, \textbf{a)}, and ZnTe, \textbf{b)}, generation. At 0~T we measure the transmitted THz waveform with and without optically pumping the silicon sample. At high magnetic field, the silicon sample is optically pumped. Data was taken at 10~K for \textbf{a)} and 83~K for \textbf{b)}.}
\label{ill:Waveforms}
\end{figure}

\begin{figure*}
\centering
\includegraphics[scale=0.55]{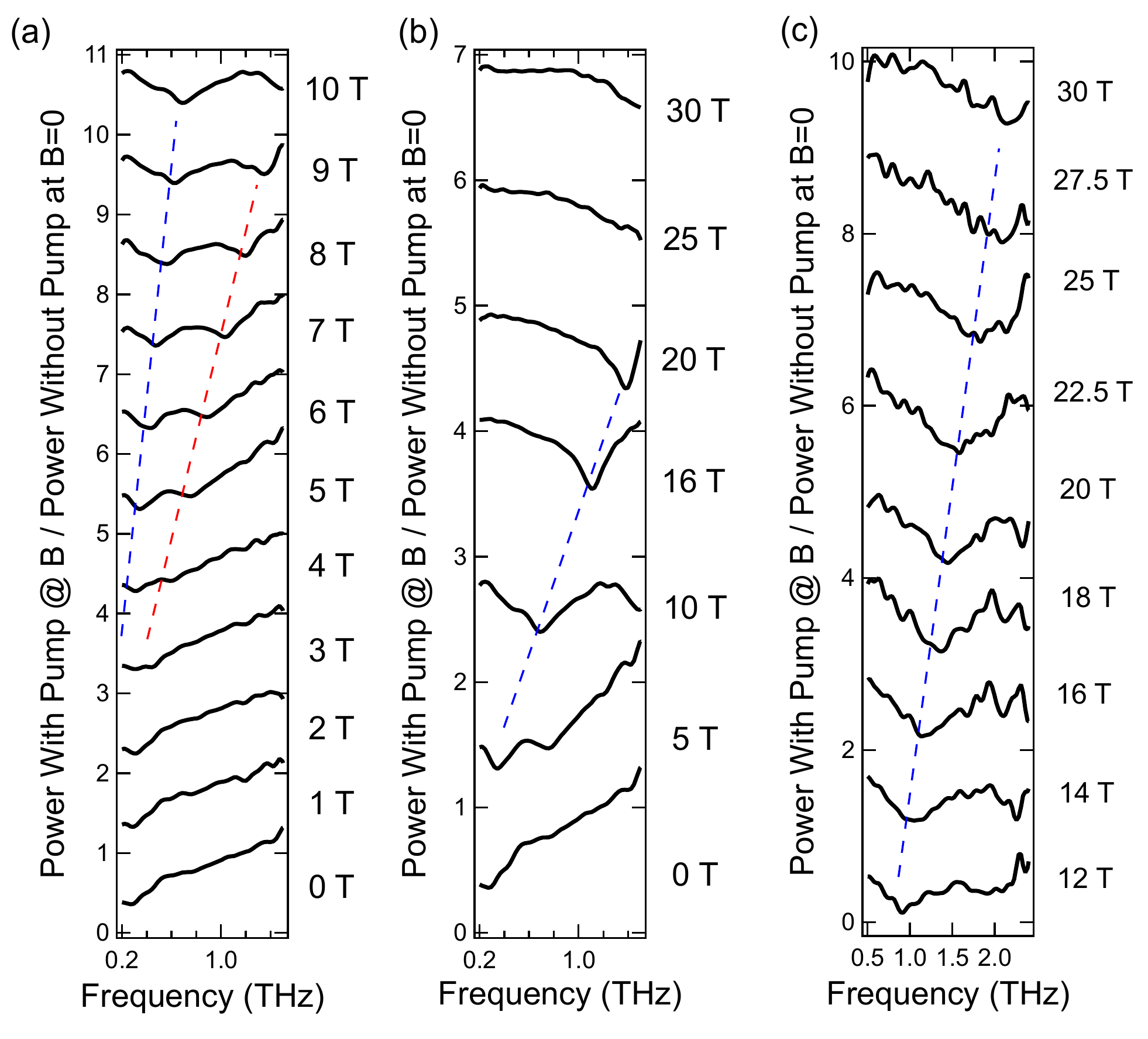}
\caption{Magnetic field dependence of the relative THz transmission for optically pumped silicon. Data shown in \textbf{a)} and \textbf{b)} were taken with LiNbO$_{3}$ generation, and the sample temperature was 10~K. Data shown in \textbf{c)} was taken with ZnTe generation, and the sample temperature was 83~K. Cyclotron resonance lines can be seen and the dashed lines are a guide to the eye for the heavier mass feature (blue) and the lighter mass feature (red). All traces are offset linearly with respect to the incremental increase in magnetic field.}
\label{ill:Transmission}
\end{figure*}

For measurements at high magnetic fields, we synchronize the magnetic field pulse and the triggering of the camera acquisition with the 1~kHz signal of the laser such that the THz radiation that passes through the sample at the peak of the magnetic field affects the polarization of the gate beam image that we measure. One image is acquired at the peak for each magnetic field pulse; see the single-shot THz trace in Fig.~\ref{ill:PulseDemo}. For the data shown in the next Section, 4 images were taken with 4 magnetic field pulses for each peak magnetic field strength. Then, we subtract a reference measurement using 100 averages and extract the THz waveform as described above. It should be noted that the term `single-shot' is somewhat of a misnomer since at minimum 2 laser pulses are required for extracting the THz electric field because a reference image with the THz beam blocked must be subtracted from the image of the gate beam with the THz incident on the detection crystal in order to eliminate artifacts and/or interference patterns in the image due to the imperfections of the optical components in the gate-beam path, including the echelon optic that are not related to the effect of the THz electric field on the polarization of the gate beam.

\section{Results and Discussion}
We performed OPTP measurements with the developed single-shot system using $\sim$10~mW of the fundamental laser output beam unfocused onto the intrinsic Si sample and a pump/probe time delay of 100~ps for both the LiNbO$_{3}$ as well as the ZnTe generation cases. Upon optical excitation, electrons from the valence band are excited above the bandgap  into the conduction band leaving holes behind. After a 100~ps delay, some of the THz radiation is absorbed by the free carriers. In an applied magnetic field, the carriers will move in circular orbits due to the Lorentz force. The cyclotron frequency, $\omega_{c} = eB/m^{*}$, is determined by $e$, the electronic charge, $B$, the applied magnetic field, and $m^*$, the effective mass.  Figure~\ref{ill:Waveforms} shows the measured waveforms after passing through the Si sample upon optical pumping at different magnetic fields, as well as the THz waveform at 0~T with no optical pump for comparison. The SNR is sufficient to clearly see a change in the electric field in the time-domain with applied magnetic field.

\begin{figure}[h]
\centering
\includegraphics[scale=0.6]{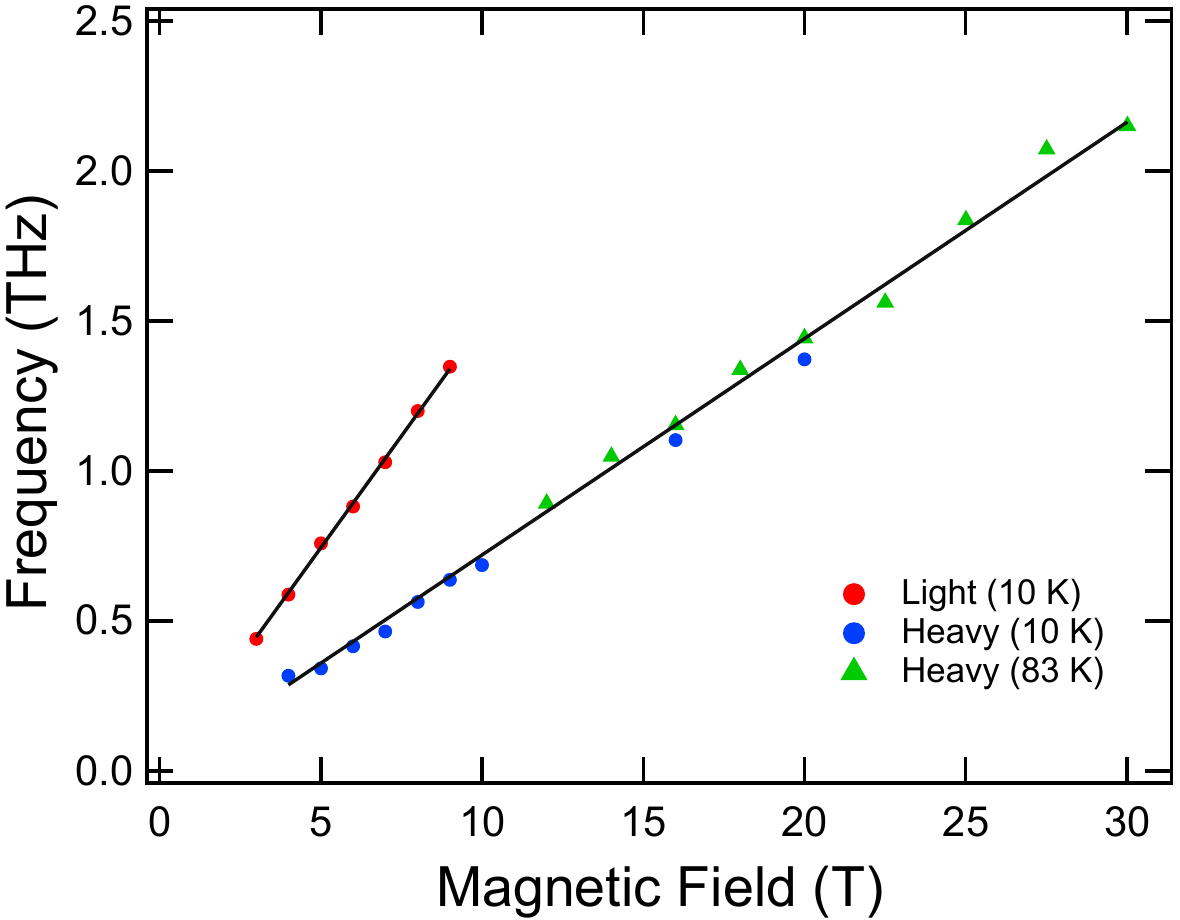}
\caption{Cyclotron resonance center frequency vs. magnetic field. Results from data taken at 10~K and 83~K with both generation schemes is combined to identify two features with frequency linear with applied magnetic field.}
\label{ill:FanDiagram}
\end{figure}

Figure~\ref{ill:Transmission} shows the relative THz transmission versus magnetic field. The relative transmission calculation uses the Fourier transform of the 0~T data without optical pumping as the reference for the data shown. At 0~T, a Drude-like response is observed in the frequency domain, where free carrier absorption increases with decreasing frequency. With increasing the magnetic field to 10~T, two dips become clearly evident in the relative transmission corresponding to cyclotron resonance absorption of the photoexcited carriers. With further increase in magnetic field above 10~T, the cyclotron resonance feature corresponding to the lower-mass carrier leaves the bandwidth of the LiNbO$_{3}$ generated THz radiation, while the feature corresponding to the higher-mass carrier remains clearly visible up to 20~T with only the tail of the feature remaining at 25~T and 30~T. Using ZnTe instead for THz generation allowed us to track the heavier mass feature up to 30~T.

After fitting the center frequency of the cyclotron resonance vs. magnetic field (Fig.~\ref{ill:FanDiagram}) with a line and calculating the effective mass, we determine the carrier effective mass, $m^{*}=0.19m_0$ for the lighter carrier and $m^{*}=0.39m_0$ for the heavier feature, where $m_0 = 9.11 \times 10^{-31}$\,kg. The lower-mass value matches very well with literature value for one electron effective mass, whereas the higher-mass value is slightly below the literature value~\cite{Dresselhaus55,Dexter56} for the heavier electron when the magnetic field is in the [100] orientation of Si.

\section{Conclusion}
We have demonstrated optical-pump/THz-probe measurements in bulk, intrinsic silicon in magnetic fields up to 30~T by developing a single-shot THz-TDS system around a minicoil pulsed magnet. The single-shot measurement faithfully reproduces the THz electric field when compared to the commonly used step scan technique. To our knowledge, these results mark the highest magnetic field reported for THz-TDS measurements. Unlike rapid scanning techniques based on ECOPS or ASOPS, this single-shot technique could potentially be used to perform THz-TDS measurements in ultrahigh magnetic field where typical magnetic field pulse durations are on the order of $\mu$s where the variation of the magnetic field during the measurement of the THz pulse would remain negligible. For ms duration pulses, the efficiency of the data taking process could be improved by an order of magnitude relative to these measurements with the use of a high frame rate camera operating at $\sim$1,000~frames per second, the same repetition rate as our 1~kHz laser. The magnetic field dependence could be taken with a single pulse instead of achieving one magnetic field data point at the peak of the magnetic field. This improvement is vitally important for the purpose of varying other parameters in addition to the magnetic field such as the sample temperature, optical-pump/THz-probe time delay, or optical pump power, as the majority of the time performing these measurements is spent waiting for the magnet coil to cool between magnet shots.

\section*{Funding}
This work was supported by the National Science Foundation through Grant No.\ DMR-1310138.  I.K.\ and J.T.\ acknowledge support from KAKENHI (Nos.\ 15K13378 and 16H04001). 

\end{document}